\documentclass{hch}
\usepackage{graphicx}

\newcommand{\be}{\begin{equation}}
\newcommand{\ee}{\end{equation}}
\newcommand{\bea}{\begin{eqnarray}}
\newcommand{\eea}{\end{eqnarray}}
\newcommand{\gsim}{\mathrel{\mathop{\kern 0pt \rlap
  {\raise.2ex\hbox{$>$}}} \lower.9ex\hbox{\kern-.190em $\sim$}}}

\def\bbox{{\,\lower0.9pt\vbox{\hrule \hbox{\vrule height 0.2 cm
\hskip 0.2 cm \vrule height 0.2 cm}\hrule}\,}}
\newcommand{\dsl}{\pa \kern-0.5em /}

\newcommand{\T}{\Theta}

\catcode`@=11 \@addtoreset{equation}{section} \catcode `@=12
\renewcommand{\theequation}{\thesection.\arabic{equation}}

\def\* {&=&}

\def\N{{\rm N}}
\newcommand{\nc}{\newcommand}
\newcommand{\rnc}{\renewcommand}

\makeatletter \rnc{\theequation}{\thesection.\arabic{equation}}
\@addtoreset{equation}{section} \makeatother

\nc{\fig}[3]{
\begin{figure}
\centerline{\epsfxsize=#1\epsfbox{#2.eps}} \caption{#3.
\label{#2}}
\end{figure}
}

\def\CK{{\cal K}}

\def\a{\alpha}

\def\d{{\rm d}}
\def\ep{\epsilon}

\def\l{\lambda}

\def\n{\nu}

\def\G{\Gamma}
\def\T{{\rm T}}

\def\bu{\noindent $\bullet$ \,}

\begin{document}
\title{Exact Answers to
Approximate Questions \\
-- Noncommutative Dipoles, Open Wilson Lines and UV-IR Duality --}
\author{Soo-Jong Rey}
\address{ School of Physics \&
Center for Theoretical Physics, Seoul National University, Seoul
151-747 \rm KOREA}
\runningtitle{Soo-Jong Rey: Exact Answers to Approximate
Questions}
\maketitle
\begin{abstract}
In this lecture, I put forward conjectures asserting that, in all
noncommutative field theories, (1) open Wilson lines and their
descendants constitute a complete set of interpolating operators
of `noncommutative dipoles', obeying dipole relation, (2) infrared
dynamics of the noncommutative dipoles is dual to ultraviolet
dynamics of the elementary noncommutative fields, and (3) open
string field theory is a sort of noncommutative field theory,
whose open Wilson lines are interpolating operators for closed
strings. I substantiate these conjectures by various intuitive
arguments and explicit computations of one- and two-loop Feynman
diagrammatics.
\end{abstract}
\section{Introduction and Conjectures}
The most salient feature of noncommutative field theories, as
opposed to the conventional commutative field theories, is that
physical excitations are described by a `noncommutative dipole'
--- weakly interacting, nonlocal object. One can visualize them as
follows. Denote their center-of-mass momentum and dipole moment as
${\bf k}$ and ${\bf \ell}$, respectively. According to the
`dipole' picture, originally developed in \cite{dipole} and more
recently reiterated in \cite{susskind}, the two quantities are
related each other: \bea {\bf \ell}^a = \theta^{ab} {\bf k}_b.
\label{relation} \eea Here, $\theta^{ab}$ denotes the
noncommutativity parameter \footnote{ Here, $\left\{ \cdot
\right\}$ refers to the Moyal commutator, defined in terms of the
$\star$-product: \bea \left\{ A(x_1) B(x_2) \right\}_\star := \exp
\left( {i \over 2} \partial_1 \wedge \partial_2 \right) A(x_1)
B(x_2) \qquad {\rm where} \qquad
\partial_1 \wedge \partial_2 := \theta^{ab} \partial_1^a \partial_2^b.
\nonumber \eea } : \bea \left\{ {\bf x}^a, {\bf x}^b
\right\}_\star = i \theta^{ab}. \label{nc} \eea Evidently, in the
commutative limit, $\theta^{ab} \rightarrow 0$, the dipole shrinks
in size and reduces to pointlike excitations.

Operationally, a noncommutative field theory is defined by an
action functional of putative {\sl elementary} fields,
collectively denoted as $\Phi$. Elementary quanta are
created/annihilated by none other than the elementary fields,
$\Phi$. Apart from radiative corrections at higher-orders, in
weak-coupling perturbation theory, they constitute a complete set
of {\sl pointlike} excitation spectrum. On the other hand, the
above argument implies that, in a generic noncommutative field
theory, in addition to point-like excitations, there ought to be
dipole-like excitations as well. An immediate and interesting
question is \sl `In what precise manner, does the noncommutative
field theory produce dipole-like excitations?' \rm. A related
questions are \sl `If the dipole-like excitations are present,
does that imply that the elementary field $\Phi$ is non-unitary?
If so, should one introduce a new field for each dipole-like
excitation?'. \rm I claim that all these provoking questions can
be answered, in noncommutative field theories, by introducing a
set of operators, nicknamed as \underline{open Wilson lines}
(OWLs) \cite{rey1, rey2}.

I found it illuminating to draw an analogy with the situation in
QED. In QED, the spectrum includes, in addition to electron and
proton quanta created/annihilated by respective fundamental
fields, bound-states such as positronium or hydrogen atom, whose
characteristic scale in wave functions or in `parton distribution
functions'is set by the Bohr radius. This does not mean, however,
that perturbative unitarity is violated, or a new field
creating/annihilating the positronium or the hydrogen atom needs
to be introduced. Rather, the bound-states are properly understood
as {\sl poles} of {\sl two-particle}-irreducible Green functions,
for which the only technical difficulty would be non-analyticity
of the fixed-order perturbation theory near the two-particle
threshold. The perturbation theory remains valid outside the
threshold region, and hence is better than the approach relying
only on dispersion relation, for which the discontinuity across
the branch cut ought to be known for {\sl all} energies. Morally
speaking, I view the noncommutative dipole as counterpart of the
QED bound-state, created/annihilated by the open Wilson line
operators.

In answering the question posed above, I put forward the following
conjectures: \hfill\break \bu 1. \underline{open Wilson lines}: In
a generic noncommutative field theory, there {\sl always} exist a
special class of composite operators,  open Wilson lines,
$W_k[\Phi]$, and their descendants, $\left(\Phi W\right)_k[\Phi]$.
\hfill\break \bu 2. \underline{noncommutative dipoles}: The open
Wilson lines and their descendants constitute a complete set of
interpolating operators for creating/annihilating the
`noncommutative dipoles', obeying the dipole relation,
Eq.(\ref{relation}). \hfill\break \bu 3. \underline{UV-IR
duality}: Infrared dynamics of the noncommutative dipoles, and
hence the open Wilson lines, $W_k[C]$, is dual to ultraviolet
dynamics of the elementary fields, $\Phi$'s.  \hfill\break \bu 4.
\underline{Closed Strings from Open Strings}: Extended to string
field theories, open Wilson lines made out of open string field
are interpolating operators for closed strings, both on and
off-shell. \hfill\break I would like to motivate these conjectures
from the following considerations. Once one recalls the Weyl-Moyal
correspondence, the conjecture 1 is rather transparent: the open
Wilson lines in Moyal formulation correspond, in Weyl formulation,
to the familiar Wilson loops, viz. `master fields' in the large $N
\sim {\rm Pf} \theta$ limit. The well-known fact that Wilson loops
form a complete set of gauge-invariant operators in the large-$N$
gauge theory motivates the first half of the conjecture 2. That
the open Wilson lines ought to obey the dipole relation,
Eq.(\ref{relation}), will be proven in section 3. The conjecture 3
constitutes the most important feature regarding the long-distance
spectrum and dynamics of generic noncommutative field theories.
Evidently, the conjectured duality is strongly reminiscent of the
`channel' duality between open and closed strings, and implies
that the open Wilson lines are operators associated with `closed'
string-like excitations in noncommutative field theories. The
conjecture 4 asserts that the open string field theory is a sort
of noncommutative field theory and that closed strings are
describable entirely in terms of the open Wilson lines made out of
open string field.

In this lecture, I take the simplest yet interacting
noncommutative field theory, $d$-dimensional $\lambda
[\Phi^3]_\star$ theory, and substantiate the above conjectures 1 -
4. Specifically, I present relevant results from computation of
one- and two-loop effective action. I prove explicitly that the
effective action is expressible entirely in terms of {\sl scalar}
open Wilson lines, and that interaction among the open Wilson
lines is noncommutative and purely geometrical. Viewing the theory
as level-zero truncation of the Witten's open string field theory,
I also argue that the scalar open Wilson lines act precisely as
the closed string field.
%
%%%%%%%%%%%%%%%%%%%%%%%%%%%%%%%%%%%%%%%%%%%%%%%%%%%%%%%%%%%%%%%%%%%%%%%%%%%
\section{Flying Noncommutative Dipole}
%%%%%%%%%%%%%%%%%%%%%%%%%%%%%%%%%%%%%%%%%%%%%%%%%%%%%%%%%%%%%%%%%%%%%%%%%%%
I begin with a physical situation from which intuitive picture of
the noncommutative dipole is developable: \underline{Mott exciton
in a strong magnetic field}. As is well-known, in a strong
magnetic field, low-energy excitation of electrons and holes is
projected to the lowest Landau level so that, in the
quasi-particle's Hamiltonian, the kinetic energy is negligible
compared to the residual potential energy such as Coulomb
interaction energy. An immediate question is whether the situation
repeats for charge-neutral excitons, viz. bound-state consisting
of an equal number of electrons and holes. Following the
pioneering works \cite{dipole}, I now prove that low-energy
excitation of the Mott exciton consists only of rigid
translational motion in the plane perpendicular to the magnetic
field.

In the non-relativistic limit, the Hamiltonian of a charge-neutral
exciton in the background of a uniform electric and magnetic
field, ${\bf E}$ and ${\bf B}$, is given by \bea
 H = {\left( {\bf p}_1 + e{\bf A}({\bf r}_1) \right)^2 \over 2m}
  + {\left( {\bf p}_2 - e{\bf A}({\bf r}_2) \right)^2 \over 2m}
  + e {\bf E} \cdot ({\bf r}_1 - {\bf r}_2 )
  + V(\vert {\bf r}_1 - {\bf r}_2 \vert).
  \nonumber
\eea
The velocity operator of the electron and the hole is given by
${\bf v}_{1,2} = \partial H \slash \partial {\bf p}_{1,2}$, and
obeys the operator equations of motion: \bea m {d {\bf v}_1 \over
dt} = - e {\bf E} + e {\bf B} \wedge {\bf v}_1 \qquad {\rm and}
\qquad m {d {\bf v}_2 \over dt} = + e {\bf E} - e {\bf B} \wedge
{\bf v}_2. \nonumber \eea
One readily finds that the total momentum of the excition, ${\bf
P} := m ({\bf v}_1 + {\bf v}_2) - e {\bf B} \wedge ({\bf r}_1 -
{\bf r}_2) $ is conserved: $d{\bf P} / d t = [H, {\bf P}] = 0$.
Take the symmetric gauge ${\bf A}({\bf r}) = {1 \over 2} {\bf B}
\wedge {\bf r}$. The conserved total momentum is then given by
\bea {\bf P} = ({\bf p}_1 + {\bf p}_2) - {e \over 2} {\bf B}
\wedge ({\bf r}_1 - {\bf r}_2). \nonumber \eea
In terms of the center-of-mass coordinate, ${\bf R}= ({\bf r}_1 +
{\bf r}_2)/2$ and the relative coordinate, ${\bf \ell}:= ({\bf
r}_1 - {\bf r}_2)$, the exciton wave-function is given by
\bea \Psi_{\bf P}({\bf r}_1, {\bf r}_2) = \exp \left( i {\bf R}
\cdot {\bf P} + {i e \over 2} {\bf R} \wedge \Delta {\bf x}
\right) \psi_{\rm P} ({\bf \ell}), \nonumber
\eea
and, after straightforward algebra, one obtains the total
Hamiltonian as
\bea H = {1 \over B^2} {\bf P} \cdot {\bf E} \wedge {\bf B} - {1
\over 2} M {{\bf E}^2 \over {\bf B}^2} + H_{\rm rel}, \qquad \quad
(M=2m) \nonumber
\eea
To make contact with noncommutative field theories more
transparent, introduce noncommutativity parameter $\theta^{ab} =
\left( {\bf F}^{-1} \right)^{ab}$, and inverse metric $G^{ab} =
(-\theta^2)^{ab}$. One then readily derives the operator relations
for the exciton center-of-mass velocity ${\bf V}$: \bea {\bf V}^a
= {\partial H \over \partial {\bf P}} = \theta^{ab} {\bf E}_b,
\nonumber \eea and for the exciton electric dipole moment ${\bf d}
\equiv e {\bf \ell}$: \bea {\bf d}^a \equiv {\partial H \over
\partial {\bf E}_a} = M G^{ab} {\bf E}_b + \theta^{ab} {\bf P}_b.
\nonumber \eea
These are precisely the noncommutative dipole relation,
Eq.(\ref{relation}).

Moreover, the relative Hamiltonian $H_{\rm rel}$ turns out
identical with the standard Landau-level problem for a charged
particle (with reduced mass). Thus, in the strong magnetic field
limit, approximating the lowest Landau level wavefunction by Dirac
delta-function, one obtains the excitation wavefunction as: \bea
\Psi_{\bf P} ({\bf R}, {\bf r}) \sim \exp \left( i{\bf R} \cdot
{\bf P} + {i e \over 2} {\bf R} \wedge {\bf \ell} \right) \delta
\left({\bf r} - {\bf \ell} \right) . \label{wavefunction} \eea
The wave-function Eq.(\ref{wavefunction})  proves that the
low-energy dynamics of the Mott exciton comprises of rigid
translation, whose dipole moment is proportional to the the
center-of-mass momentum.
%%%%%%%%%%%%%%%%%%%%%%%%%%%%%%%%%%%%%%%%%%%%%%%%%%%%%%%%%%%%%%%%%%%%%%%%%%%
\section{Open Wilson Lines: How and Why?}
%%%%%%%%%%%%%%%%%%%%%%%%%%%%%%%%%%%%%%%%%%%%%%%%%%%%%%%%%%%%%%%%%%%%%%%%%%%
\subsection{Open Wilson Lines}
%%%%%%%%%%%%%%%%%%%%%%%%%%%%%%%%%%%%%%%%%%%%%%%%%%%%%%%%%%%%%%%%%%%%%%%%%%%
How, in a given noncommutative field theory, are the open Wilson
lines defined, and what are they for? I claim that the answer lies
ultimately to the observation alluded above: noncommutative
dipoles are present generically as part of theory's low-energy
excitation. In noncommutative {\sl gauge} theories, the answer
also has to do with gauge-invariant, physical observables, so I
will begin with this case first. In \cite{rey1,rey2, gross}, it
has been shown that (part of) gauge orbit is equivalent to the
translation along the noncommutative directions. For example, in
noncommutative U(1) gauge theory, the gauge potential ${\bf
A}_\mu(x)$ and the neutral scalar field ${\Phi}(x)$, both of which
give rise to `noncommutative dipoles', transform in `adjoint'
representation: \bea \delta_\epsilon {\bf A}_\mu (x) &=& i \int
{\d^2 {\bf k} \over (2 \pi)^2} \, \widetilde{\epsilon}({\bf k})
\Big[ \Big( {\bf A}_\mu ({\bf x} + \theta \cdot {\bf k}) - {\bf
A}_\mu ({\bf x} - \theta \cdot {\bf k}) \Big) + i {\bf k}_\mu
\Big]
e^{ i {\bf k} \cdot {\bf x}} \nonumber \\
\delta_\epsilon {\Phi}(x) &=& i \int {\d^2 {\bf k} \over (2
\pi)^2} \, \widetilde{\epsilon}({\bf k}) \Big[ {\Phi} ({\bf x} +
\theta \cdot {\bf k}) - {\Phi} ({\bf x} - \theta \cdot {\bf k})
\Big] e^{ i {\bf k} \cdot {\bf x}}, \label{gaugetransf} \eea
where the infinitesimal gauge transformation parameter is denoted
as \bea \epsilon({\bf x}) = \int {\d^2 {\bf k} \over (2 \pi)^2}
e^{ i {\bf k} \cdot {\bf x}} \, \widetilde{\epsilon}({\bf k}).
\nonumber \eea Because of the peculiarity, for fields transforming
in `adjoint' representations under the noncommutative gauge group,
there is {\sl no} physical observables local in configuration
space. The only physical observable one can construct turn out
local in momentum space, and is referred as the `open Wilson line'
operators \cite{kawai, rey1, rey2, gross} and their descendants
defined on an open contour $C$: \bea W_{\bf k} [{\bf A}] &=& {\cal
P}_{\rm t} \int {\d^2} {\bf x} \, \exp_\star \left( i \int_0^1 \d
t \, \dot{\bf y} \cdot {\bf A}
                  (x + {\bf y}) \right)
\star e^{ i {\bf k} \cdot {\bf x}},
\label{openwilson} \\
\left( {\cal O} W \right)_{\bf k} [{\bf A}] &=& {\cal P}_{\rm t}
\int \d^2 {\bf x} \left[ \int\limits_0^1 \d t {\cal O}(t)
       \exp_\star \left( i \int_0^1 \d t \, \dot{\bf y} \cdot {\bf A}
                         (x + {\bf y}) \right)
\right] \star e^{ i {\bf k} \cdot {\bf x}}. \nonumber \eea Here,
the $\star$-product is defined with respect to the base point
${\bf x}$ of the open contour $C$. Despite being defined over an
open contour, the operator is gauge-invariant {\sl provided} the
momentum ${\bf k}$ is related to the geodesic distance ${\bf y}(1)
- {\bf y}(0) := \Delta {\bf x}$ precisely by the `dipole
relation', Eq.(\ref{relation}). In other words, in noncommutative
gauge theory, the open Wilson lines (physical observables) are
noncommutative dipoles, obeying the dipole relation
Eq.(\ref{relation}) as an immediate consequence of the gauge
invariance!

The open Wilson lines are actually ubiquitous and are present in
generic noncommutative field theories, in which neither gauge
invariance nor gauge field is present. This is because, as I have
convinced you already, the dipole relation Eq.(\ref{relation})
ought to be a universal relation, applicable for {\sl any}
theories defined over noncommuative spacetime. In \cite{oneloop,
oneloop2, oneloop3}, I have shown that the scalar open Wilson line
operators $W_{\bf k}[\Phi]$ and descendants $(\Phi^n W)_{\rm k}
[\Phi]$ are given by: \bea W_{\bf k}[\Phi] &:=& {\cal P}_{\rm t}
\!\! \int \!\! \d^2 {\bf x}\, \exp \left( i g \int_0^1 \d t \vert
\dot{\bf y}(t) \vert \, \Phi(x + {\bf y}(t))  \right) \star e^{ i
{\bf k} \cdot {\bf x}}
\nonumber \\
(\Phi^n W)_{\bf k}[\Phi] &:=& \left( - i {\partial \over \partial
g} \right)^n W_{\bf k}[\Phi] \qquad (n = 1, 2, 3, \cdots),
\label{scalarwilson}
\eea
where
$\lambda$ is an appropriate coupling parameter.

Can one show that the dipole relation Eq.(\ref{relation}) is
satisfied for scalar fields, wherein no gauge invariance is
present? Consider the following set of operators, so-called Parisi
operators \cite{parisi}: \bea {\cal O}_n (x_1, \cdots, x_n; {\bf
k}) = \int d^2 {\bf z} \, \Phi_1 (x_1 + {\bf z}) \star \Phi_2 (x_2
+ {\bf z}) \star \cdots \star \Phi_n (x_n + {\bf z}) \star e^{ i
{\bf k} \cdot {\bf x}}, \nonumber \eea viz. Fourier-transform of a
string of elementary scalar fields, $\Phi_k(x)$ $(k = 1, 2,
\cdots)$. Take the one-point function: \bea G_1 (x, {\bf k}) :=
\left< {\cal O}_2 (x, {\bf k}) \right>  = \left< \int d^2{\bf z}
\, \Phi({\bf z}) \star\Phi(x + {\bf z}) \star e^{ i {\bf k} \cdot
{\bf x}} \right>. \nonumber \eea In terms of Fourier decomposition
of the scalar field: \bea \Phi(x) = \int {d^2 {\bf k} \over (2
\pi)^2} \, e^{ i {\bf k} \cdot {\bf x}} \widetilde{\Phi}({\bf k}),
\nonumber \eea I obtain that \bea G_1 (x, {\bf k}) = \int {d^2
{\bf l} \over (2 \pi)^2} \widetilde{\Phi}({\bf l})
\widetilde{\Phi}(-{\bf l} + {\bf k}) \exp\left[ i {\bf l} \cdot
\left(x + {1 \over 2} \theta \cdot {\bf k} \right)\right].
\nonumber \eea Consider `wave-packet' of the scalar particle,
$\Phi({\bf z}) = \Phi_0 \delta^{(2)} ({\bf z})$ and $\Phi(x +{\bf
z}) = \Phi_0 \delta^{(2)} (x + {\bf z})$ so that
$\widetilde{\Phi}({\bf l}) = \Phi_0 \exp(i {\bf l} \cdot x)$. From
the above equation, I then find that \bea G_1 (x, {\bf k}) =
\Phi_0^2 \delta^{(2)} \left( x +{1 \over 2} \theta \cdot {\bf k}
\right). \nonumber \eea Thus, I find that the stationary point of
the correlator is given by $\Delta x^a \, \sim \, \theta^{ab} {\bf
k}_b, $ and hence precisely by the `dipole relation',
Eq.(\ref{relation}).

I also claim that the open Wilson lines are a sort of `master
fields'. According to the Weyl-Moyal correspondence, generic
noncommutative fields, be they the gauge field ${\bf A}$ or the
scalar field $\Phi$, are interpretable as $(N \times N)$
matrix-valued fields at $N \rightarrow \infty$ limit, living only
on commutative directions (if there is any). Hence, from the
latter formulation, one can construct Wilson loop operators as the
large-$N$ master fields: \bea W_{\bf k}[\widehat{A}] = {\rm Tr}
\exp \left( i {\bf k} \cdot \widehat{\bf A} \right) \qquad {\rm
or} \qquad W_k [\widehat{\Phi}] = {\rm Tr} \exp \left( i k
\widehat{\Phi} \right). \nonumber \eea In fact, one can readily
show that, once expanded around the noncommutative space, these
Wilson loop operators turn into the aforementioned open Wilson
lines.
%%%%%%%%%%%%%%%%%%%%%%%%%%%%%%%%%%%%%%%%%%%%%%%%%%%%%%%%%%%%%%%%%%%%%%%%%%%
\subsection{Generalized Star Products}
%%%%%%%%%%%%%%%%%%%%%%%%%%%%%%%%%%%%%%%%%%%%%%%%%%%%%%%%%%%%%%%%%%%%%%%%%%%
Computationally, the open Wilson lines originate from resummation
of so-called generalized $\star$-products. As such, I will first
indicate how the generalized $\star$-products are inherent to the
defintion of the open Wilson lines.

Begin with the gauge open Wilson lines. For a straight contour,
expanding Eq.(\ref{openwilson}) in successive powers of the gauge
field ${\bf A}_\mu$, it was observed \cite{mehenwise, liu3} that
generalized $\star$-product, $\star_n$, a structure discovered
first in \cite{garousi, liu}, emerge: \bea W_{\bf k} [C] = \int
\d^2 {\bf x} \, \left[ 1 - \left( \partial \wedge {\bf A} \right)
+ {1 \over 2!} \left(\partial \wedge {\bf A} \right)^2_{\star_2} +
\cdots \right] \star e^{ i {\bf k} \cdot {\bf x}}.
\label{wilsonline2} \eea

The generalized $\star_n$ product exhibits different algebraic
structures from Moyal's $\star$-product. For instance, the first
two, $\star_2, \star_3$ defined as
\bea \left[A(x_1) B(x_2) \right]_{\star_2} &:=& { \sin \left({1
\over 2} \partial_1 \wedge
\partial_2 \right) \over {1 \over 2}
\partial_1 \wedge \partial_2} A(x_1) B(x_2)
\nonumber \\
\left[ A(x_1) B(x_2) C(x_3) \right]_{\star_3} &:=& \left[ {\sin
\left( {1 \over 2} \partial_2 \wedge \partial_3 \right) \over {1
\over 2}(\partial_1 + \partial_2) \wedge \partial_3 } {\sin
\left({1 \over 2} \partial_1 \wedge (\partial_2 +\partial_3)
\right) \over {1 \over 2} \partial_1 \wedge (\partial_2
+\partial_3)} +(1 \leftrightarrow 2) \right] \nonumber \\
&& \times A(x_1) B(x_2) C(x_3) \nonumber \eea show that the
$\star_n$'s are commutative but non-associative. Despite these
distinguishing features, I interpret Moyal's $\star$ product more
fundamental than the generalized $\star_n$ products. The open
Wilson line is defined in terms of path-ordered $\star$-product,
and its expansion in powers of the gauge potential involves the
generalized $\star_n$-product at each $n$-th order. As such,
complicated $\star_n$ products arise upon expansion in powers of
the gauge potential, and are attributible to dipole nature of the
open Wilson line and the gauge invariance therein -- each term in
Eq.(\ref{wilsonline2}) is {\sl not} gauge invariant, as the gauge
transformation Eq.(\ref{gaugetransf}) mixes terms involving
different $\star_n$'s. Indeed, the generalized $\star_n$ products
are not arbitrary but obey recursive identities: \bea i
\left[\partial_x A \wedge
\partial_x B \right]_{\star_2}
&=& \left\{ A, B \right\}_{\star} \nonumber \\
i \partial_x \wedge \left[ A \, B \partial_x C \right]_{\star_3}
&=& A \star_2 \left\{ B, C \right\}_\star + B \star_2 \left\{A, C
\, \right\}_\star. \nonumber \eea
These identities are crucial for ensuring gauge invariance of the
power-series expanded open Wilson line operator,
Eq.(\ref{wilsonline2}).

I can show readily that the same sort of generalized
$\star$-products also show up in the scalar open Wilson lines.
Consider a simplified form of the scalar open Wilson line with an
insertion of a local operator ${\cal O}$ at a location ${\bf R}$
on the Wilson line contour: \bea ({\cal O}_{\bf R} W)_{\rm k}
[\Phi] := {\cal P}_{\rm t} \int \d^2 {\bf x} \, {\cal O}(x + {\bf
R}) \star \exp \left( i g \int_0^1 \d t \vert \dot{\bf y}(t) \vert
\Phi(x + {\bf y}(t) ) \right) \star e^{ i {\bf k} \cdot {\bf x}}.
\nonumber \eea Take, for simplicity, a {\sl straight} Wilson line:
\bea {\bf y}(t) = {\bf L} t \qquad {\rm where} \qquad {\bf L}^a =
\theta^{ab} {\bf k}_b := \left( \theta \cdot {\bf k} \right)^a,
\quad L := \vert {\bf L} \vert, \nonumber \eea corresponding to a
{\sl uniform} distribution of the momentum ${\bf k}$ along the
Wilson line. As the path-ordering progresses to the right with
increasing $t$, power-series expansion in $g L \Phi$ yields:
\bea ({\cal O}_{\bf R} W)_{\rm k}[\Phi] &=& \int \d^2 {\bf x} e^{
i {\bf k} \cdot {\bf x}} \star \left[ {\cal O}({\bf x} + {\bf R})
\right.
\nonumber \\
&+& igL \int\limits_0^1 \d t {\cal O}(x + {\bf R}) \star \Phi(x +
{\bf L} t)
\nonumber \\
&+& (igL)^2 \int\limits_0^1 \d t_1 \int_{t_1}^1 \d t_2 {\cal
O}({\bf x} + {\bf R}) \star \Phi(x + {\bf L} t_1) \star \Phi(x +
{\bf L} t_2 )
\nonumber \\
&+& \left. \cdots \quad \right]. \nonumber \eea
I can evaluate each term, for instance, by Fourier-transforming
${\cal O}$ and $\Phi$'s: \bea {\cal O}(x) = \int {\d^2 {\bf k}
\over (2 \pi)^2} \widetilde{O}({\bf k}) \, { T}_{\bf k}, \quad
\Phi(x) = \int {\d^2 {\bf k} \over (2 \pi)^2}
\widetilde{\Phi}({\bf k}) \, {T}_{\bf k} \quad {\rm where} \quad
{T}_{\bf k} = e^{ i {\bf k} \cdot {\bf x}}, \nonumber \eea taking
the $\star$-products of the translation generators: ${T}_{\bf k}
\star {T}_{\bf l} = e^{ {i \over 2} {\bf k} \wedge {\bf l}} \,
{T}_{{\bf k} + {\bf l}}$, and then evaluating the parametric $t_1,
t_2, \cdots$ integrals. Fourier-transforming back to the
configuration space, after straightforward calculations, I obtain
 \bea ({\cal O}_{\bf R}
W)_{\rm k}[\Phi] &=& \int \d^2 {\bf x} e^{ i {\bf k} \cdot {\bf
x}} {\cal O}(x + {\bf R})
\label{taylor} \\
&+& igL  \int \d^2 {\bf x} \, e^{ i {\bf k} \cdot {\bf x}} \left[
{\cal O}(x + {\bf R}) \Phi(x)\right]_{\star_2}
\nonumber \\
&+& {1 \over 2!} (i gL)^2  \int \d^2 {\bf x} \, e^{ i {\bf k}
\cdot {\bf x}} \left[{\cal O}(x + {\bf R}) \Phi(x) \Phi(x)
\right]_{\star_3} + \cdots. \nonumber \eea
I then observe that the products involved, $\star_2, \star_3,
\cdots$, are precisely the {\sl same} generalized $\star_n$
products as those appeared prominently in the {\sl gauge} open
Wilson line operators.
%%%%%%%%%%%%%%%%%%%%%%%%%%%%%%%%%%%%%%%%%%%%%%%%%%%%%%%%%%%%%%%%%%%%%%%%%%%
\section{Free and Interacting OWLs}
%%%%%%%%%%%%%%%%%%%%%%%%%%%%%%%%%%%%%%%%%%%%%%%%%%%%%%%%%%%%%%%%%%%%%%%%%%%
To identify degrees of freedom associated with the open Wilson
lines and to understand their spectrum and interaction, consider
the $\lambda [\Phi^3]_\star$-theory, and study the effective
action. The noncommutative Feynman rules of
$\lambda[\Phi^3]_\star$-theory are summarized, in the background
field method,  by the following generating functional:
\bea Z[\Phi_0] = Z_0[\Phi_0] \int {\cal D} \varphi \exp \left( -
\int \d^d x \left[ {1 \over 2} \varphi (x) {\cal D}_{\Phi_0}
\varphi(x) + {\lambda \over 3!} \varphi^3(x) \right]_\star
\right). \label{rule} \eea
Here, ${\cal D}_{\Phi_0} = ( - \partial_x^2 +m^2 + \lambda
\Phi_0(x))$, and $\Phi_0(x)$ and $\varphi(x)$ refer to the
background and the fluctuation parts of the scalar field, $\Phi$,
respectively. Because of the noncommutativity of the
$\star$-product, interactions are classifiable into planar and
nonplanar ones. I focus on so-called nonplanar part of the one-
and two-loop Feynman diagrams, and, as I am interested primarily
in dynamics at long distance, on the low-energy and large
noncommutativity limit:
\bea {p \over m} = {\cal O}\left(\epsilon^{+1}\right), \quad m^2
\theta^{ab} = {\cal O}\left(\epsilon^{-2}\right), \quad {\lambda
\Phi \over m^2} = {\cal O}(\epsilon^{+1}) \quad {\rm as} \quad
\epsilon \rightarrow 0^{+}. \label{lelimit} \eea
%%%%%%%%%%%%%%%%%%%%%%%%%%%%%%%%%%%%%%%%%%%%%%%%%%%%%%%%%%%%%%%%%%%%%%%%%%%
\subsection{Free OWLs}
%%%%%%%%%%%%%%%%%%%%%%%%%%%%%%%%%%%%%%%%%%%%%%%%%%%%%%%%%%%%%%%%%%%%%%%%%%%
Begin with effective action at one loop. The nonplanar part of the
one-particle-irreducible N-point Green function is given by (see
Fig. 1)
\bea \G_\N \left(\{p_i\}, \{q_j\} \right) &=& \hbar
\left(-\frac{\l}{2}\right)^\N \sum_{\N_1+\N_2=\N} {C_{\{N\}} \over
(4 \pi)^{d/2}} \int_0^{\infty} \frac{\d\T}{\T} \T^{-{d \over
2}+\N} \nonumber \\
&&\times \, \exp\left[-m^2 \T-\frac{\ell^2}{4\T} \right]
J_{\N_1}(\ell) J_{\N_2}(-\ell). \nonumber \eea Here, I have
denoted an N-dependent combinatoric factor as $C_{\{\N\}}$,
divided $\N$ external momenta into two groups: $\{p_1, \cdots
p_{\N_1}\}$ and $\{ q_1, \cdots, q_{\N_2} \}$, and defined $k =
\sum_{i=1}^{\N_1} p_i = - \sum_{i=1}^{\N_2} q_i$, and $\ell :=
\theta \cdot k$ (consistent with the noncommutative dipole
relation). I have also defined $J_\N(\ell)$ by \bea J_{\N_1}
(\ell,\{p_i\}) := \int_{-1/2}^{1/2} \d \tau_1 \cdots \d \tau_\N \,
\exp\left[-i\sum_{i=1}^{\N_1} \tau_i p_i\cdot \ell - \frac{i}{2}
\sum_{i<j=1}^{\N_1}\ep(\tau_{ij}) p_i\wedge p_j \right], \nonumber
\eea
where $\tau_{ij} \equiv (\tau_i-\tau_j)$, and similarly $J_{\N_2}
(- \ell, \{q_j\})$. They are precisely the momentum-space kernel
of the generalized $*_\N$ product.
\begin{figure}[hbtp]
    \centering
    \includegraphics[width=5.truecm]{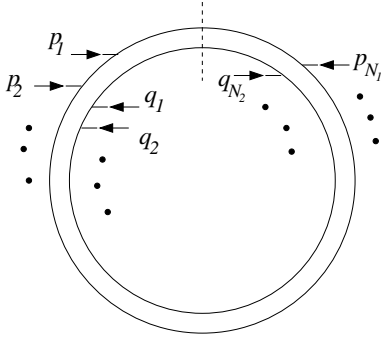}
    \caption[]{One-loop Feynman diagram for N-point one-particle-irreducible
Green function. The circumference of the vacuum diagram has a
length T, while the relative position of each external line from
the reference point (dashed vertical) is denoted $\tau_i$'s.}
    \label{oneloop}
\end{figure}

In the limit Eq.(\ref{lelimit}), the $\T$-moduli integral is
evaluated accurately via the saddle-point approximation.
Evidently, $\T = |l|/2m$ at the saddle point. Then, the integral
is evaluated as
\bea \G_\N [\{p_i\}, \{q_j\}] &=& \hbar \left(2 \pi
\frac{|\ell|}{m}\right)^{-{d \over 2}} \left( {2 \pi \over m \vert
\ell \vert} \right)^{1/2} e^{-m|\ell|} \nonumber \\
&\times& \sum_{\N_1+ \N_2= \N} C_{\{\N\}} \left\{
g^{\N_1}|l|^{\N_1} J_{\N_1}(\ell) \right\} \left\{
g^{\N_2}|l|^{\N_2} J_{\N_2}(-\ell) \right\}, \nonumber \eea
where $g \equiv - \l/4m$. The factorized expression permits
exponentiation of the double-sum over $\N_1, \N_2$.  Indeed,
summing over $\N = \N_1 + \N_2$, taking carefully into account of
the combinatorial factors $C_\N$, I find \cite{oneloop, oneloop2}
\bea \G &=& \frac{\hbar}{2} \int \frac{\d^d \ell}{(2\pi)^d}
W(\ell) \, \CK_{-d}(|\ell|) \, W(-\ell), \nonumber \eea
where
\bea \CK_{-d} (\vert \ell \vert) &=& \left( 2 \pi {\vert \ell
\vert \over m } \right)^{-{d \over 2}} \left( {2 \pi \over m \vert
\ell \vert } \right) e^{-m\vert \ell \vert},
\nonumber \\
W(\ell) &=& \sum_{\N=0}^\infty \int {\d^d p_1 \over (2 \pi)^d}
\cdots \int {\d^d p_\N \over (2 \pi)^d} (2 \pi)^d \delta^d \left(
p_1 +\cdots + p_\N - \ell\right)
\nonumber \\
&&\times {1 \over \N!} (-g \vert \ell \vert )^\N \left[ \Phi(p_1)
\cdots \Phi(p_\N) \cdot J_\N (\{p_i\}, \ell) \right]. \nonumber
\eea
In the last step, I utilized the result of the previous subsection
that open Wilson lines are expandable in power-series of
$\star_\N$ products.

\begin{figure}[hbtp]
    \centering
    \includegraphics[width=5.truecm]{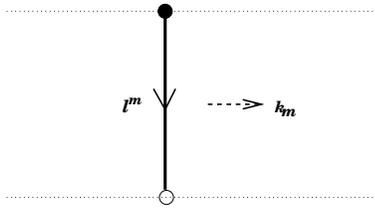}
    \caption[]{Spacetime view of open Wilson line propagation. The
noncommutativity is turned on the plane spanned by the two
vectors, $\ell^m$ and $k_m$.}
    \label{propagator}
\end{figure}

The $\N_1 = \N_2 = 1$ term was computed previously in
\cite{seiberg}, where infrared singular behavior of the result was
interpreted as manifestation of the UV-IR duality. What remained
not understood in the work of \cite{seiberg} was understanding
reason behind the UV-IR duality. What I have shown above is that
this is answerable by summing over N, viz. computing the full
effective action, and that the UV-IR duality originates from
noncommutative dipole degrees of freedom inherent in any
noncommutative field theory. In fact, the one-loop effective
action is expressible schematically as
\bea \Gamma_2 [W] = {1 \over 2} {\rm Tr}_{{\cal H}_{\rm dipole}}
\Big( \widehat{W} \,\cdot \, {\bf K}_2 \, \cdot \, \widehat{W}
\Big), \qquad \qquad \qquad {\bf K}_2 := \widehat{\cal K}_{-{d
\over 2}}. \nonumber \eea
Here, I have used the Weyl-Moyal correspondence and expressed the
open Wilson lines and kernel as operators defined on one-dipole
Hilbert space ${\cal H}_{\rm dipole}$. Remarkably, the action
takes strikingly the same form as the quadratic part of {\sl
matter} sector in Witten's cubic open string field theory!
%%%%%%%%%%%%%%%%%%%%%%%%%%%%%%%%%%%%%%%%%%%%%%%%%%%%%%%%%%%%%%%%%%%%%%%%%%%
\subsection{Interacting OWLs}
\begin{figure}[hbtp]
    \centering
    \includegraphics[width=5.truecm]{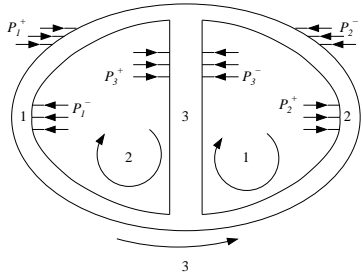}
    \caption[]{Two-loop Feynman diagram for N-point Green function. The external
lines and the vacuum diagram boundaries are labelled {\em dual}
each other. The external momenta $p_i^{(a)}$, Feynman-Schwinger
moduli parameters $\tau_i^{(a)}$, and Moyal's phase-factor signs
$\n_i^{(a)}$ are ordered from top to bottom. With this ordering
convention, the $\tau's$ range over $0 < \tau_\N < \cdots < \tau_1
< \T$ for each connected side of the three internal propagators.}
    \label{setup}
\end{figure}

%%%%%%%%%%%%%%%%%%%%%%%%%%%%%%%%%%%%%%%%%%%%%%%%%%%%%%%%%%%%%%%%%%%%%%%%%%%
I next compute the nonplanar part of the two-loop effective action
and show that it is expressible as cubic interaction of the open
Wilson lines. Begin with the two-loop nonplanar contribution to
the $N$-point one-particle-irreducible Green functions
\cite{twoloop}. The Feynman diagram under consideration is
depicted in Fig.\ref{setup}. Begin with constructing a {\sl
planar} two-loop vacuum diagram \footnote{At two loop and beyond,
vacuum diagrams are classifiable into a planar diagram and the
rest, nonplanar diagrams. If the number of twist insertion is
zero, the vacuum diagram is referred as planar. All other vacuum
diagrams, with at least one insertion of the twist, are nonplanar
ones. At one loop, by default, the vacuum diagram is planar.}
constructed by joining three internal propagators via two
$\lambda[\Phi^3]_\star$ interaction vertices. Denote the internal
propagators in double-lines and label them as $a=$ 1, 2, and 3 in
Fig.(\ref{setup}). Moduli parameters $\T_1, \T_2, \T_3$ refer to
the Feynman-Schwinger parameters of the three internal
propagators, and range over the moduli space, ${\cal M}_{\rm
2-loop} = [0, \infty) \otimes [0, \infty) \otimes [0, \infty)$. We
then affix N external lines (background fields), distributed among
the three internal propagators as $\N_a$ ($a=1,2,3$) so that
$(\N_1 + \N_2 + \N_3) = \N$. Each group of external lines are
further classifiable into those affixed from the inner and the
outer boundaries. Sum over all possible insertion of the external
lines is provided by integration over the moduli parameters
$\tau_i^{(a)}$'s over $[0, \T_a]$, and $\tau_{ij}^{(a)}$ refers to
$(\tau_i^{(a)} - \tau_j^{(a)})$.

Momenta of external lines attached at $a$-th internal propagator
are labelled as $\{ p_i^{(a)}\}$, where $a = 1,2,3$ and $i = 1, 2,
\cdots, \N_a$. I introduced total momentum injected on $a$-th
internal propagator as:
\bea P_a^\pm \equiv \sum_{i=1}^{\N_a} \frac{1 \pm \n_i^{(a)} }{2}
p_i^{(a)} \qquad {\rm and} \qquad P_a \equiv P_a^+ + P_a^- ~,
%\label{ftmm}
\nonumber \eea
where $\pm$ refers to the inner and the outer boundaries,
respectively, and $\n_i^{(a)}$ takes $\pm1$ depending on whether
the insertion is made from the `left' or the `right' side. I have
also introduced the total momentum inserted on each {\em
worldsheet boundary} via $k_{a+2} = P^+_{a} + P^-_{a+1}$.
Again, taking the limit Eq.(\ref{lelimit}) and consuming a lengthy
computation, I have obtained:
\bea\label{GPN} \G_\N \left(\{p_a^{(a)}\} \right) = {\hbar^2
\lambda^2 \over 24} \left(-{\lambda \over 2} \right)^\N
\sum_{\{\N_a\}=0}^\N \sum_{\{\nu\}} (2\pi)^d \delta
\left(\sum_{a=1}^3\sum_{n=1}^{\N_a}p_n^{(a)} \right) \,C_{\{\N\}}
\, \Gamma_{\nu }^{(\{\N_a\})}. \nonumber \eea
Here, $C_{\{\N\}}$ denotes a combinatoric factor, and
\bea
\begin{array}{rcl}
\Gamma_{\sigma , \nu}^{(\N_1, \N_2, \N_3)} &=& \displaystyle
\int\limits_0^\infty {\d \T_1 \d \T_2 \d \T_3 \over (4 \pi)^d} \,
e^{F(\T_1, \T_2, \T_3)} \Delta^{\frac{d}{2}}(\T) \left(
\prod_{a=1}^3 \int\limits_0^{\T_a} \prod_{i=1}^{\N_a} \d
\tau^{(a)}_i \right)
\nonumber\\
&\times& \prod_{a=1}^{3} \left( \sum_{\N_a^+}
\frac{\T_a^{\N_a^+}}{\N_a^+!} J_{\N_a^+}(+\a_a) \right) \left(
\sum_{\N_a^-} \frac{\T_a^{\N_a^-}}{\N_a^-!} J_{\N_a^-}(-\a_a)
\right)
\end{array}
\label{owlegg} \eea in which \bea \exp F(\T_1, \T_2, \T_3) = \exp
\left[ - m^2 (\T_1 + \T_2 + \T_3) - {\Delta \over 4} (\T_1
\ell_1^2 + T_2 \ell_2^2 + \T_3 \ell_3^2 ) \right], \nonumber \eea
\bea \Delta^{-1}(\T) := \T_1 \T_2 + \T_2 \T_3 + \T_3 \T_1,
\label{Delta} \eea
\bea \widehat{J}_{(\N^+_a, \N^-_{a+1})} \left( - \alpha_a,
\alpha_{a+1} \right) = \exp \left[ {i \over 2} t_a t_{a+1} k_a
\wedge k_{a+1} \right] \widetilde{J}_{\N^+_a}
\widetilde{J}_{\N^-_{a+1}} \nonumber \eea
and $\widetilde{J}$'s are precisely the same as the one-loop
$\star_\N$ kernel. I have also introduced the following shorthand
notations:
\bea t_a &=& \sqrt{\Delta} \T_a, \qquad (t_1 t_2 + t_2 t_3 + t_3
t_1 = 1)
\nonumber \\
\alpha_1 &=& t_1(t_2 \ell_2 - t_3 \ell_3), \quad \alpha_2 = t_2
(t_3 \ell_3 - t_1 \ell_1),\quad \alpha_3 = t_3 (t_1 \ell_1 - t_2
\ell_2). \nonumber \eea
Geometrically, for any given nonnegative values of $\{t_a\}$,
$\{\alpha_a\}$ split the triangle formed by $\{\ell_a\}$ into
three pieces, viz., $\ell_1 = \alpha_3 - \alpha_2$, $\ell_2 =
\alpha_1 - \alpha_3$, $\ell_3 = \alpha_2 - \alpha_1$.

The $\T$-moduli integrals are computable by saddle-point
conditions. ${\partial F} \slash {\partial \T_a} = 0$ yields
\bea
 \Delta^{-1} = \frac{L^2}{4m^2}
\quad {\rm and} \quad
 L \equiv | t_1 l_1 - t_2 l_2 |
 = |t_2 l_2 - t_3 l_3 | = |t_3 l_3 - t_1 l_1 .
\label{anglesad} \eea
They determine the ``size" of the moduli and their relative
``angles". Geometrically, the condition Eq.(\ref{anglesad})
demands that the angle between a pair of $\a$'s is $2\pi/3$.
Moreover, the value of $F (\{\T_a\})$ at the saddle point also has
a simple geometric description:$ F ({\rm saddle}) = - m \left( |
\alpha_1 |
 + | \alpha_2 | +  |\alpha_3| \right)_{\rm saddle}$.
 The crucial point is that, at the saddle point, I now have
\bea \T_a = \Delta^{-1/2} t_a = \frac{L}{2m} t_a
    = \frac{ |\alpha_a | }{2m} ~ ,
\label{wow} \eea
Plugging Eq.(\ref{wow}) into Eq.(\ref{owlegg}), each factor sums
up to an open Wilson line
\bea \sum_{\N_a^+} \left( - \frac{\lambda }{2} \right)^{\N_a^+}
  \frac{\T_a^{\N_a^+}}{\N_a^+!} J_{\N_a^+}(\a_a)
= \sum_{\N_a^+} \left( - \frac{\lambda} {4 m}
   | \alpha_a | \right)^{\N_a^+} J_{\N_a^+}(\a_a) ~ ,
\nonumber \eea
just like the one-loop case, except that the Wilson line contour
is now `snapped'! I finally obtain the two-loop effective action
as \cite{pants}
\bea \label{finalfantasy} \G[\ell] &=& {1 \over 3} \lambda^2
\hbar^2 \int \frac{\d^d k_1}{(2\pi)^d} \cdots \frac{\d^d
k_3}{(2\pi)^d} \delta^{(d)}(k_1+k_2+k_3)
\left(\frac{2m}{L}\right)^{d-3} \left({\delta \T \over \T
}\right)^3
\nonumber \\
&\times& \exp\Big[-m(|\a_1|+|\a_2|+|\a_3|)\Big] \exp \left( -{i
\over 2} \sum_{a=1}^3 \alpha_a \wedge \alpha_{a+1} \right)
\nonumber \\
&\times& [\widehat{W}(\a_1, -\a_2)] [\widehat{W}(\a_2,-\a_3)]
[\widehat{W}(\a_3,-\a_1)]. \nonumber \eea

\begin{figure}[hbtp]
    \centering
    \includegraphics[width=5.truecm]{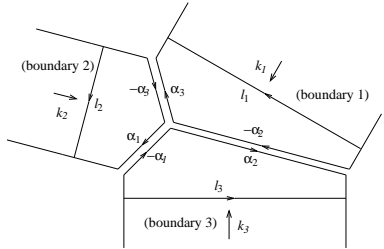}
    \caption[]{Spacetime view of interaction among three open Wilson lines. At
asymptotic region, the dipoles are described by straight open
Wilson lines, while, at interaction region, the dipoles interact
locally pairwise by `snapping' the open Wilson lines.}
    \label{}
\end{figure}
Again, the two-loop effective action is expressible schematically
as:
\bea \Gamma_3[W] = {\lambda_{\rm c} \over 3} {\rm Tr}_{{\cal
H}_{\rm dipole}} {\bf K}_3 \, \Big( \widehat{W} \, \widehat{\star}
\, \widehat{W} \, \widehat{\star} \, \widehat{W} \Big) \qquad {\rm
where} \qquad \lambda_{\rm c} = (\lambda/2)^2, \label{cubic} \eea
where ${\bf K}_3$ represents a weight-factor over one-dipole
Hilbert space ${\cal H}_{\rm dipole}$, and the
$\widehat{\star}$-product refers to a newly emergent
noncommutativity in the algebra of open Wilson lines. I note that
the two-loop effective action, once expressed in terms of the
snapped open Wilson lines, is remarkably similar to `half-string'
picture of the cubic interaction in Witten's open string field
theory, again restricted to the matter sector, with a minor
difference that location of the triple interaction point is
determined {\sl dynamically} by the energy-momentum carried by the
asymptotic dipoles. Note also that `soft-dilaton theorem' is
obeyed --- the interaction strength $\lambda_{\rm c}$ is
proportional to {\sl square} of the $\Phi$-field coupling
parameter, $\lambda$.

%%%%%%%%%%%%%%%%%%%%%%%%%%%%%%%%%%%%%%%%%%%%%%%%%%%%%%%%%%%%%%%%%%%%%%%%%%%
\section{Closed Strings out of Open Strings}
%%%%%%%%%%%%%%%%%%%%%%%%%%%%%%%%%%%%%%%%%%%%%%%%%%%%%%%%%%%%%%%%%%%%%%%%%%%
The last conjecture of mine concerns extrapolation of the
observations made in noncommutative field theories to string
theories: {\sl closed strings} are created/annihilated by `open
Wilson lines' made out of open string fields \cite{sjrey}.

The starting point would be that, at level-zero truncation and in
the background of nonzero two-form potential $B$, Witten's cubic
open string field theory is approximated by a noncommutative field
theory of the open string tachyon with a cubic interaction. Expand
the tachyon potential around local minimum. The resulting theory
reduces precisely to the $\lambda[\Phi^3]_\star$-theory I have
discussed at length already, but, quite importantly, with the
mass-squared $m^2 \rightarrow \infty$. According to Sen's
conjecture, excitation around the tachyon potential minimum ought
to correspond to that of a closed string only and none of the open
string. A point to be explored is `can one understand Sen's
conjecture even with zero-level truncation, viz. a noncommutative
$\lambda[\Phi^3]_\star$ scalar field theory? I now claim that the
answer to this question is affirmatively yes.

\subsection{Open String as Miniature Dipole}
The starting point is the observation that open strings in a
constant B-field backgrounds are noncommutative dipoles, albeit of
miniature size. In conformal gauge, an open string worldsheet
action is
\bea S_{\rm open} = {1 \over 2 \ell_{\rm st}^2} \int_\Sigma \left[
\left(G_{mn}\delta^{\alpha \beta} + B_{mn}\epsilon^{\alpha
\beta}\right)
\partial_\alpha X^m \partial_\beta X^n \right] + \int_{\partial_{\rm L} \Sigma - \partial_{\rm R} \Sigma}
A_m(X) \dot X^m , \nonumber \eea
where $\partial_{\rm L, R} \Sigma$ refers to left or right
boundary of the worldsheet $\Sigma$. The $B$-field is locally
exact, and can be gauged away via U(1) transformation: $B_2
\rightarrow B_2 + d \Lambda_1$, $ A_1 \rightarrow A_1 - \ell_{\rm
st}^{-2} \Lambda_1$. This results in
\bea
S_{\rm open} = {1 \over 2 \ell_{\rm st}^2} \int_\Sigma  G_{mn}
\left( \dot X^m \dot X^n - X'^m X'^n \right) + \int_{\partial_{\rm
L} \Sigma - \partial_{\rm R} \Sigma} A_m (X) \dot X^m. \nonumber
\eea
I now latticize the open string by two points, separated by
worldsheet length $2 \ell_{\rm st}$. Let $X(0,\tau) = X_{\rm
L}(\tau)$ and $X(2 \ell_{\rm st}, \tau) = X_{\rm R}(\tau)$. The
open string action is then approximatable as
\bea
S_{\rm open} &\rightarrow& \int d\tau \left[ {1 \over 2}m
\left(\dot X_{\rm L}^2 + \dot X_{\rm R}^2 \right) - {1 \over 2}
m\omega^2
(X_{\rm L} - X_{\rm R})^2 \right] \nonumber \\
&+&  \int d\tau \left[ q B_{mn} \left(\dot X_{\rm L}^m X_{\rm L}^n
- \dot X_{\rm R}^m X_{\rm R}^n \right) \right], \nonumber \eea
where $m = \omega = 1/\ell_{\rm st}$ and $q = 1/\ell_{\rm st}^2$.
Alternatively, in terms of center-of-mass and relative
coordinates, ${\bf R} = (X_{\rm L} + X_{\rm R})/2$ and $\ell =
(X_{\rm L} - X_{\rm R})$,
\bea
S_{\rm open} = \int d\tau \left[ {1 \over 2} M\dot{\bf R}^2 + {1
\over 2} \mu \dot\ell^2 - {1 \over 2} \mu \omega^2 \ell^2 + {q
\over 2} B_{mn} \left(\dot{\bf R}^m \ell^n - \dot\ell^m {\bf R}^n
\right) \right], \nonumber
\eea
where $M = 2m$, $\mu = m/2$. In either form, it shows that the
two-point latticized open string is literally identical to the
Mott exciton or the noncommutative dipole. From the action, I also
infer the boundary conditions:
\bea m \omega^2 (X_{\rm L} - X_{\rm R})_m + q F_{mn}(X_{\rm L})
\dot X_{\rm L}^n &=& 0, \nonumber \\
m \omega^2 (X_{\rm R} - X_{\rm L} )_m - q F_{mn} (X_{\rm R}) \dot
X_{\rm R}^n &=& 0. \label{bc} \eea
I now quantize the open string, viz. first-quantize the two point
particles inside the dipole. In doing so, the boundary conditions
Eq.(\ref{bc}) needs to be imposed as constraints. The dipole
coordinates then obey exactly the same commutation relations as
those obeyed by the Mott exciton:
\bea
\left[X_{\rm L}^m, X_{\rm L}^n \right] = + i \theta^{mn}, \qquad
\left[X_{\rm R}^m, X_{\rm R}^n \right] = - i \theta^{mn}, \qquad
\left[X_{\rm L}^m, X_{\rm R}^n \right] = 0, \nonumber
\eea
or, equivalently, in the notation adopted for the Mott exciton,
\bea
\left[ {\bf R}^m, \ell^n \right] = i \theta^{mn}, \qquad \left[
{\bf R}^m, {\bf R}^n \right] = 0 = \left[\ell^m, \ell^n \right].
\nonumber
\eea
These commutation relations indicate that the open string is a
noncommutative dipole, obeying the dipole relation:
one-dipole
Hilbert space is simply a tensor product of two one-particle
Hilbert spaces: ${\cal H}_{\rm dipole} = {\cal H}_{\rm L} \otimes
{\cal H}_{\rm R}$. In light of the latticization employed,
one-dipole Hilbert space should be identified with first-quantized
string Hilbert space.

\subsection{Witten's $\star_{\rm w}$-product is Moyal's
$\star_{\rm m}$-product} Utilizing the dipole picture of
latticized open string, I now argue that the $\star$-product
defining Witten's open string theory is identifiable as Moyal's
$\star$-product. Actually, in the absence of constant B-field
background, the isomorphism holds for string oscillator modes, but
not for zero mode. Once B-field is turned on, the isomorphism hold
exactly including the zero mode. This is the reason why we started
with Witten's open string field theory in B-field background.

Schematically, Witten's $\star_{\rm w}$ product is defined on
string Hilbert space ${\cal H}_{\rm L} \otimes {\cal H}_{\rm R}$
by
\bea \left( \vert x_{\rm L} \rangle \otimes \vert x_{\rm R}
\rangle \right) \star_{\rm w} \left( \vert y_{\rm L} \rangle
\otimes \vert y_{\rm R} \rangle \right) \longrightarrow \langle
x_{\rm R} \vert y_{\rm L} \rangle \, \left( \vert x_{\rm L}
\rangle \otimes \vert y_{\rm R} \rangle \right). \label{wittens}
\eea
Because of the dipole relation, I advocate the viewpoint treating
the dipole configuration space $(X_{\rm L}, X_{\rm R})$ as a
one-particle phase-space $({\bf R}, {\bf P})$ associated with
dipole's center-of-mass, where ${\bf P} = \theta^{-1} \cdot \ell$.
Then, via Weyl-Moyal correspondence, Moyal's $\star_{\rm m}$
product in $({\bf R}, {\bf P})$ space ought to be equivalent to
matrix product in $(X_{\rm L}, X_{\rm R})$ space. It then follows
that Moyal's $\star_{\rm m}$ product equals to \sl Fourier
transform \rm of Witten's $\star_{\rm w}$ product.

\begin{figure}[hbtp]
    \centering
    \includegraphics[width=3.5truecm]{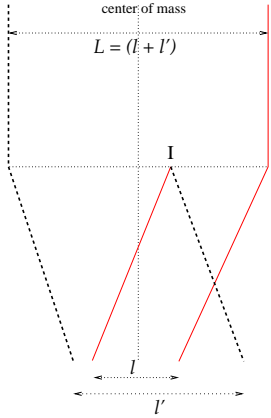}
    \caption[]{Interaction of two miniature dipoles in support of
equivalence between Moyal's $\star_{\rm m}$ product and Witten's
$\star_{\rm w}$ product. Interaction point $I$ can be anywhere and
should be summed over. } \label{equivalence}
\end{figure}

Explicitly, start with Moyal's $\star_{\rm m}$-product in $({\bf
R}, {\bf P})$ space:
\bea
[A \star_{\rm m} B]({\bf R}, {\bf P}) = A ({\bf R}, {\bf P})
\exp\left[{i \over 2} \left(\overleftarrow{\partial_{\bf R}} \cdot
\overrightarrow{\partial_{\bf P}} - \overleftarrow{\partial_{\bf
P}} \cdot \overrightarrow{\partial_{\bf R}} \right) \right] B({\bf
R}, {\bf P}). \label{moyals}
\eea
I next "Fourier transform" with respect to ${\bf P}$ and express
all in terms of dipole's relative distance $\ell$'s:
\bea
A({\bf R}, {\bf P}) = \int d \ell e^{ - i {\bf P} \cdot \ell}
\widetilde{A}({\bf R}, \ell) \quad {\rm and} \quad B({\bf R}, {\bf
P}) = \int d \ell e^{ - i {\bf P} \cdot \ell} \widetilde{B}({\bf
R}, \ell).
 \nonumber
\eea
Substituting so, Eq.(\ref{moyals}) is considerably simplified
after a change of variables: $L = (\ell + \ell')$, $L' = (\ell' -
\ell)$. Fourier transforming back the whole expression in
Eq.(\ref{moyals}) with respect to ${\bf P}$, I obtain
\cite{barsrey}:
\bea [\widetilde{A \star_{\rm m} B}] ({\bf R}, L) =
\int\limits_{-\infty}^{+\infty} {d L' \over 2} \widetilde{A}({\bf
R} + L/2, {\bf R} + L'/2) \widetilde{B} ({\bf R} + L/2, {\bf
R}-L'/2). \label{final}
\eea
The emerging picture is that a miniature dipole $\widetilde{A}$ at
center ${\bf R}$ and of length $\ell$ and another $\widetilde{B}$
at center ${\bf R}$ and  of length $\ell'$ come into contact. When
interacting, $\widetilde{A}, \widetilde{B}$ shift their centers to
${\bf R} + \ell'$ and ${\bf R} - \ell$, respectively. The final
dipole $\widetilde{A \star_{\rm m} B}$ is then centered at ${\bf
R}$ and of length $L = (\ell + \ell')$. See Fig.\ref{equivalence}.
Evidently, the dipole interaction Eq.(\ref{final}) defined via
Moyal's $\star_{\rm m}$ product yields is algebraically equivalent
to the string field interaction Eq.(\ref{wittens}) defined via
Witten's $\star_{\rm w}$ product.

\subsection{Closed Strings as OWLs}
Recall that I have identified the scalar field $\Phi$ with the
level-zero mode of the open string field. If I focus on low-energy
and low-momentum excitation below a fixed cutoff, $p^2 \le
\Lambda^2$, as $m^2 \rightarrow \infty$, excitation of the
$\Phi$-quanta is entirely suppressed. This is clearly counterpart
of half of Sen's conjecture: `around the tachyon potential
minimum, there is no open string excitation'.  The regime
$\Lambda^2 \ll m^2$ is also of considerable relevance to the
effective action computation in $\lambda [\Phi^3]_\star$-theory,
which I have not discussed at all so far. The point is that, in
addition to the nonplanar diagram contribution, there also exists
the planar diagram contribution to the effective action. The
planar part is actually sensitive to the UV cutoff. If I identify
the UV cutoff with the fixed cutoff $\Lambda$ and take the
conventional limit $m^2 \ll \Lambda^2$, the planar part of the
effective action yields a sort of Coleman-Weinberg type potential
(plus derivative corrections) -- viz. exponentiation of the scalar
field $\Phi$ takes place. On the other hand, if I take the
opposite limit, $\Lambda^2 \ll m^2$, I have found that the planar
diagram contribution turns remarkably into the same functional
form as the nonplanar diagram contribution, except that (some of)
the open Wilson lines carry nearly zero momentum. The point is
that, even for planar diagrams, the scalar field $\Phi$ is
exponentiated into open Wilson lines, albeit miniature ones,
\underline{provided} the cutoff condition obeys $\Lambda^2 \ll
m^2$. While quite bizzare from the standard quantum field theory
viewpoint, to one's delight, this cutoff condition is precisely
what is dictated by Witten's open string field theory!

The other half of Sen's conjecture --- closed string out of open
string tachyon vacuum --- is then readily inferred from the
results of previous sections. The open Wilson line formed out of
the tachyon field $\Phi$ is precisely the interpolating operator
creating and annihilating a closed string. The fact that open
Wilson lines are Moyal formulation counterpart of the Wilson loop
in Weyl formulation adds an another supporting evidence for this
claim \cite{sjrey}. There is one peculiar aspect, though. First of
all, the spacetime structure of the open Wilson lines is literally
open, viz. the two ends are situated at distinct points in the
target space. Moreover, the cubic interaction of the open Wilson
lines, Eq.(\ref{cubic}), involves newly emergent
$\widehat{\star}$-product. As both are the aspects inherently
associated with traditional open strings, one might feel
suspicious to my conjecture of identifying the open Wilson lines
as closed strings. I claim that a resolution can be drawn from the
well-known fact that closed string is formed by joining two ends
of open string(s). In the absence of the two-form potential,
$B_{mn} = 0$, size of the open string is characteristically of
string scale, and is too small to be probed by the level-zero
truncated tachyon field. If the two-form potential is nonzero,
$B_{mn} \ne 0$, the open string is polarized to a size much bigger
than the string scale, and behaves essentially like a rigid rod.
Because of that, joining and splitting of the two end of open
string(s) would never form a closed string. In other words, open
Wilson lines are precisely what the open strings can do the best
for forming closed strings out of themselves! Reverting the logic,
utility of turning on the B-field and hence noncommutativity for
the open string is to render closed strings as much the same as
open strings. That the open Wilson lines interaction is governed
by a newly emergent $\widehat{\star}$-product (see
Eq.(\ref{cubic}) would then constitute a nontrivial prediction of
the conjectures I put forward.

\section*{acknowledgement}
 This work was supported in part by BK-21
Initiative in Physics (SNU - Project 2), KOSEF Interdisciplinary
Research Grant 98-07-02-07-01-5 and KOSEF Leading Scientist
Program.

\end{document}